\documentclass{ws-ijmpa}
\usepackage[super,compress]{cite}
\usepackage{graphicx}
\begin{document}

\title{A global significance evaluation and expected significance estimation with additional data using simulated events}

\author{Kelly J Yi}

\address{Metea Valley High School\\
Aurora, Illinois 60563, USA\\
yikelly888@gmail.com}

\author{Leonard G Spiegel}

\address{Fermi National Accelerator Laboratory\\
Batavia, Illinois 60510, USA\\
lenny@fnal.gov}

\author{Zhen Hu}

\address{Department of Physics, Tsinghua University\\
Beijing 100084, China\\
zhenhu@tsinghua.edu.cn}

\maketitle

\begin{abstract}

In High-Energy Physics experiments it is often necessary to evaluate the global statistical significance of apparent resonances observed in invariant mass spectra. One approach to determining significance is to use simulated events to find the 
probability of a random fluctuation in the background mimicking a real signal. As a high school summer project, we demonstrate a method with Monte Carlo simulated events to 
evaluate the global significance of a potential resonance with some assumptions. This method for determining significance is general and can be applied, with appropriate modification, to other resonances.


\end{abstract}



\section{Introduction}	

There are often discoveries in various areas of research, and the first question we would like to ask is how significant it is and at what level we should believe it, which can be quantitatively evaluated by the significance of these discoveries. The standard for claiming a ``discovery'' usually requires significance greater than 5 standard deviations. The chance that background fluctuation to produce a $5\sigma$ fluctuation physically consistent with a real resonance is less than one in a million, assuming the statistics of a Gaussian distribution. Since the background fluctuation probability is so small, it is the consensus of the High-Energy Physics (HEP) field that such resonances represent physical states. (It is still important for discoveries to be confirmed by other experiments for full acceptance.)  On the other hand, if the significance of a resonance exceeds 3 standard deviations but is less than 5 standard deviations, the claim is classified as ``evidence’’ for a new state. For $3\sigma$ the probability of a random background fluctuation is slightly greater than one in a thousand, and although this is still very small, there is always some doubt that the resonance could be the result of a non-random fluctuation. Typically, there is a progression over time from an initial indication to evidence for and, finally, to the discovery of a new state. This was recently seen in the 2012 joint announcement by the ATLAS and CMS collaborations
~\footnote{The Large Hadron Collider (LHC) at CERN is the largest and highest energy accelerator in the world. Two protons beams are each accelerated to 7 TeV energy inside a 27-kilometre ring and collide with each other. ATLAS (short for ``A Toroidal LHC Apparatus'') and CMS (short for ``Compact Muon Solenoid'') are two general-purpose particle detectors at LHC. As two of the largest international high energy physics collaborations, ATLAS and CMS have achieved many important scientific results, such as the discovery of the Higgs boson in 2012, which is the last elementary particle in Standard Model (SM).} 
of the discovery a Higgs-like boson if not the actual Standard Model Higgs boson~\cite{Higgs-ATLAS,Higgs-CMS}. The discovery claim was furthered strengthened by the existence of multiple channels and by the independent corroboration of two groups. 

The significance can also be evaluated by accounting for a ``look elsewhere'' effect~\cite{look-else-where}, which takes into account that a fluctuation can appear anywhere in the search window (so the fluctuation probability is increased). However, if the local significance is very high this effect is typically ignored. In the case of the discovery of the $X(3872)$~\cite{X3872-Belle} exotic hadron
~\footnote{According to the Standard Model, the basic building blocks of all matter in the universe are quarks and leptons. Each baryon is composed of three quarks, while each meson is composed of a quark and an antiquark. Baryons and mesons are collectively called hadrons. However, when Gell-mann first proposed the schematic quarks model in 1964\cite{GellMann}, he also mentioned the possibility of exotic hadrons, quark systems composed of four or more quarks. No clear experimental evidence was found over the next 40 years, only in 2003 the discovery of a new particle $X(3872)$ by the Belle Collaboration at KEK opened the new era for exotic hadrons~\cite{X3872-Belle}. A possible interpretation of $X(3872)$ is a tetra-quark made of a charm quark, an anti-charm quark, and two other light quark-antiquark pair. } 
, for example, the local significance was more than 10 standard deviations, which means the chance that background fluctuated as large as signal is as low as $10^{-24}$ and a look elsewhere analysis was not done. In contrast, a new particle $X(4140)$ was observed with a local significance of 5.3 standard deviations by CDF~\cite{Y4140-CDF1,Y4140-CDF2}, and in this case look elsewhere effect was considered necessary, which led to a ``global'' significance of 4.3 standard deviations~\cite{Y4140-CDF1}. There are many other examples, including the discovery of penta-quarks by the LHCb experiment.~\cite{LHCb-penta}

As a high school summer research project, the objective of this study is to evaluate the global significance of a potential new particle with unknown mass and width uncertainties using simulated events. As a method demonstration, we took a posterior scenario for the mass and width range with some assumptions.

\section{Background and Signal Components}

In this exercise, the background components are modeled with two threshold functions. The signal component is described by a standard relativistic Breit-Wigner formula:

\begin{eqnarray}
\begin{array}{rcl}
  BW(m;m_0,\Gamma_0) &=& 
  \displaystyle{\frac{\sqrt{m\Gamma(m)}}{m_0^2 - m^2 - i m\Gamma(m)}}, \\[8pt]
  \Gamma(m) &=& \displaystyle{\Gamma_0} \left(\frac{q}{q_0}\right)^{2L + 1}\frac{m_0}{m}\left(B^{\prime}_L(q, q_0, d)\right)^2 , \\[8pt]
  B^\prime_L(q, q_0, d) 
  &=& \displaystyle{\frac{q^{-L}B_L(q,d)}{q_0^{-L}B_L(q_0,d)}}
  = \left(\frac{q_0}{q}\right)^L\frac{B_L(q,d)}{B_L(q_0,d)}, \\[8pt]
  B_0(q, d) &=& 1, \\[8pt]
  B_1(q, d) &=& \displaystyle{\sqrt{\frac{2z}{z+1}}},\\[8pt]
  B_2(q, d) &=& \displaystyle{\sqrt{\frac{13z^2}{(z-3)^2 + 9z}}},\\[8pt]
  B_3(q, d) &=& \displaystyle{\sqrt{\frac{277z^3}{z(z-15)^2 + 9(2z-5)^{2}}}},\\[8pt]
  B_4(q, d) &=& \displaystyle{\sqrt{\frac{12746z^4}{(z^2-45z+105)^2 + 25z(2z-21)^2}}},\\[8pt]
  z &=& (|q|d)^2, z_0 = (|q_0|d)^2, 
\end{array}
\label{eqn3}
\end{eqnarray}

where the parameter $L$ is set to be zero, which is called S-wave in particle physics and is the simplest case.

\section{Null-Hypothesis and Signal-Hypothesis Test}

The signal and background formulas are expressed as a PDF (Probability Density Function) in ROOT/RooFit\cite{root} software package. Each PDF can be used to generate simulated events according to its distribution through a function called $generate(RooArgSet(x), n)$, where $x$ the variable and $n$ is the number of events to be generated.  

First we do a log-likelihood fit to the simulated data by excluding the existence of signal PDF, but including all background PDF components in the fit. We obtain a log-likelihood value of the minimized fit --- $L0$ for the fit. This fit is called the null-hypothesis fit, which means that we assume the tested component does not exist.  Then we do another fit that includes all singal and background components, and we obtain a minimized log-likelihood value of this fit --- $L1$. This fit is called the signal-hypothesis fit. The log-likelihood value of a fit is a parameter to show how good a fit is to the data, and lower value means better fit --- typically the signal-hypothesis fit has a lower value.  The difference between $L1$ and $L0$ shows the statistical size of the effect of the tested component. One way to quantify the effect is to compute $\sqrt{2*(L0-L1)}$, which converts the likelihood difference into a significance. This significance is called the local significance. 

\section{Global Significance Investigation}
A significance can always be evaluated with simulated background-only events by looking for the appearance of large fluctuations.  Possible fluctuations can be looked at a specific position with specific width if we know these values in priori, otherwise we need to look at a specific range defined in advance for both mass and width --- such a significance is called a global significance. Things cannot be decided completely in advance in many cases, and thus comprises must always be made. Without theoretical guidance as to the mass and width of an exotic signal, though, we need to look at a wide range based on plausible values for the mass and width, so that the significance will necessarily be global.

\begin{figure}[b]
\centerline{\includegraphics[width=10cm]{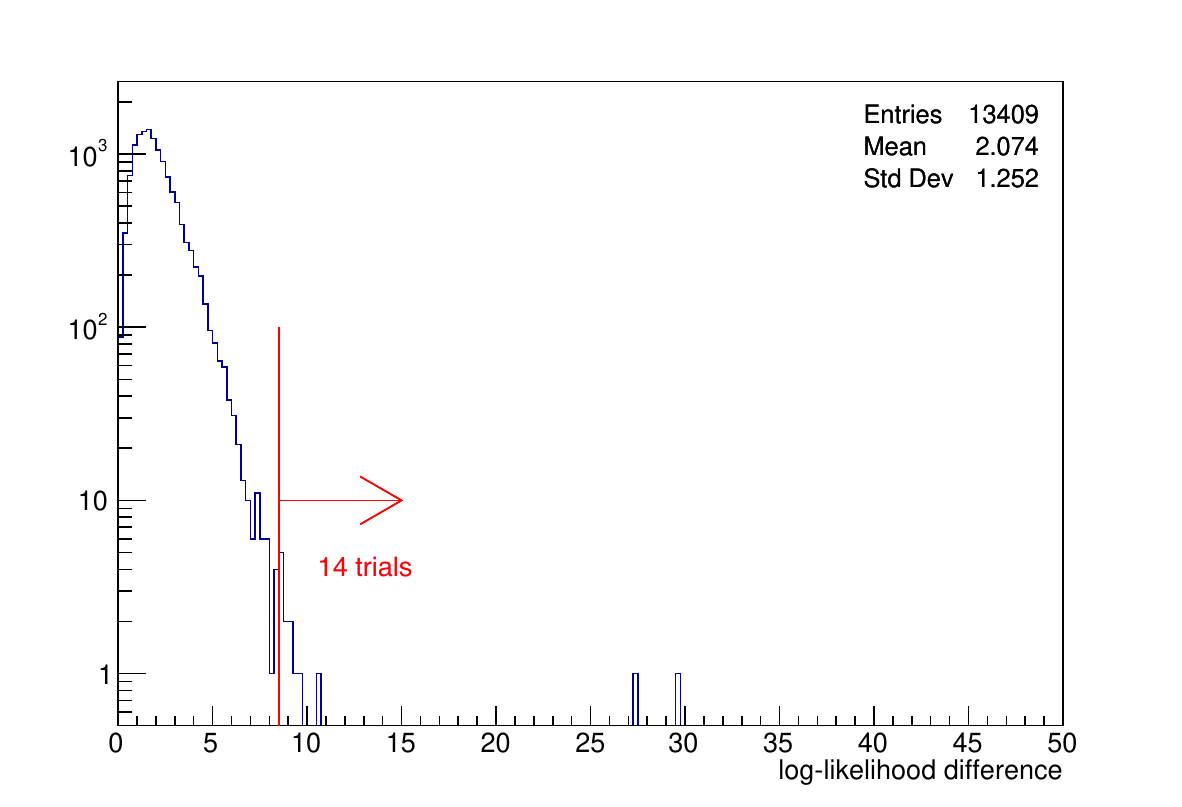}}
\caption{Log-likelihood difference, $(L0-L1)_{trial}$, of 13409 trials. The red line and arrow shows one example that 14 trials with value greater than 8.5 ($(L0-L1)_{observed}$). \label{f2}}
\end{figure}

In this study, we submitted several thousand jobs to a computing farm in order to perform many trials. In each trial, we can get a mass spectrum from simulation events generated with background-only PDF, leaving out signal PDF. Then we searched for fluctuations in the mass region and calculate 
$(L0-L1)_{trial}$ for the largest fluctuation found in each trial. 
Fig.~\ref{f2} shows $(L0-L1)_{trial}$ for these trails. We define a variable, $(L0-L1)_{observed}$, to express the value observed in real experimental data, then we count the number of the trials with $(L0-L1)_{trial} > (L0-L1)_{observed}$ in Fig.~\ref{f2}, then calculate corresponding p-value and significance. 

\begin{figure}[b]
\centerline{\includegraphics[width=10cm]{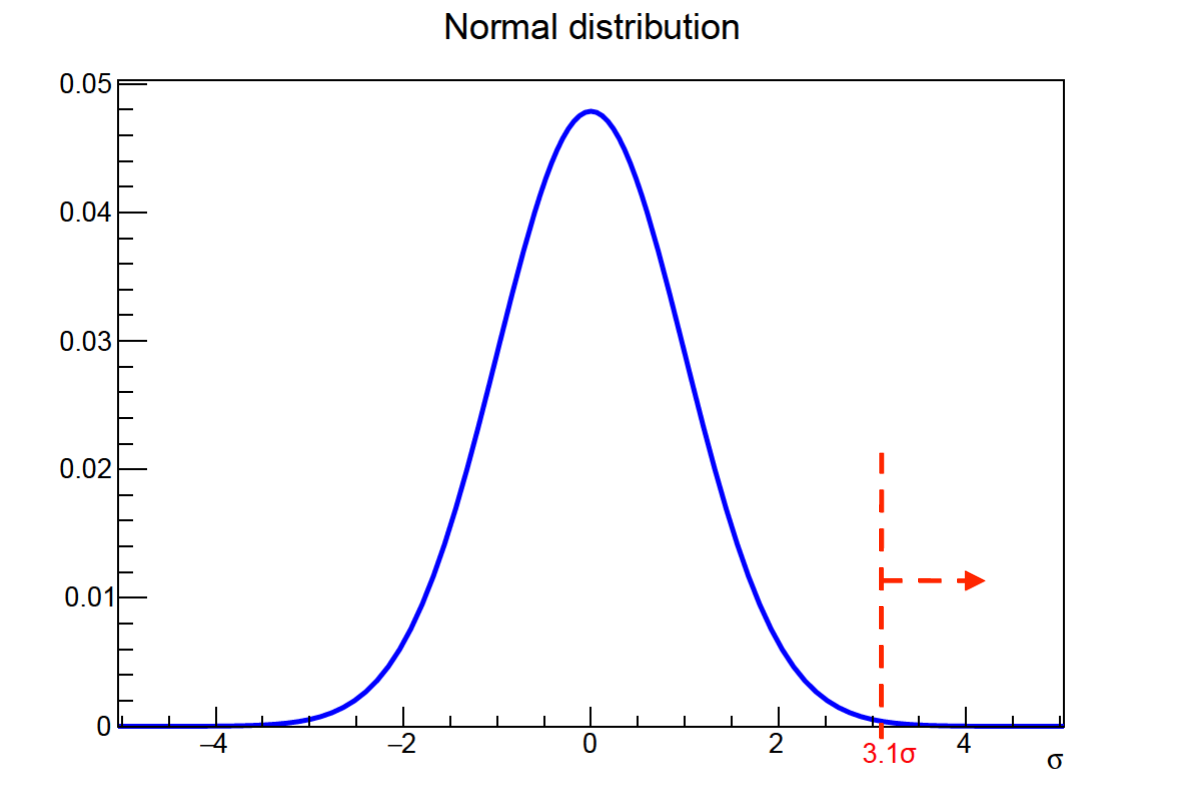}}
\caption{$3.1 \sigma$ as shown in a normal distribution. \label{f3}}
\end{figure}

For example, assuming $(L0-L1)_{observed}=8.5$, there are 14 out of the 13409 trials in Fig.\ref{f2} with $(L0-L1)_{trial}$ over 8.5, so the probability is calculated as $14/13409=0.00104$, which is called the p-value. 
This small p-value signifies a low probability of this resulting from a random fluctuation. We used the statistical tool built in RootStats\cite{root}, $RooStats::PValueToSignificance(0.00104)$, to convert the probability into significance. Assuming the probability follows that of a Gaussian distribution, our p-value of 0.00104 corresponds to the area of the integral from $3.1 \sigma$ to positive infinity for a normalized Gaussian distribution, as shown in Fig.~\ref{f3}. 

We assumed several $(L0-L1)_{observed}$, 7.75, 8.25, 8.5, 8.75, 9.0, and count the number of trials with $(L0-L1)_{trial} > (L0-L1)_{observed}$ for each case, and then calculated the corresponding p-value and significance as shown in Table~\ref{table}. 

\begin{table}
    \centering
    \begin{tabular}{cccc}
        \hline
        $(L0-L1)_{observed}$ & $N_{trials}$ with $(L0-L1) > (L0-L1)_{observed}$ & p-value & Significance \\
        \hline
        7.75 & 25 & 0.00186 & 2.9 \\
        8.25 & 18 & 0.00134 & 3.0 \\
        8.5 & 14 & 0.00104 & 3.1 \\
        8.75 & 9 & 0.00067 & 3.2 \\
        9.0 & 7 & 0.00052 & 3.3 \\
        \hline
    \end{tabular}
    \caption{$(L0-L1)$ threshold   }
    \label{table}
\end{table}

\section{Expected Significance Estimation with Additional Data}
To find out if a newly found signal is real, we often check it in additional independent data sample. Normally this is called an update.  However,  fluctuations do exists in the additional data sample, and a typical question we would like to ask before doing an update is how much additional data we need to combine to reach 5$\sigma$ significance. For instance, using the above example, for a new signal with a local significance of 4.1$\sigma$, people normally use the following formula to estimate the expected significance in the updated analysis: expected significance = $4.1\sigma \times \sqrt{NF}$, where $NF$ is the fraction of the combined data size compared to the original data size. The expected significance reaches 5$\sigma$ when $NF$ is 1.5 in this example, that is to say, the 4.1$\sigma$ new signal can reach 5$\sigma$ if the original data is combined with $50\%$ additional data.  This is only true in a very ideal situation, which in reality depends on many factors such as background shapes, background increasing fraction, etc..   

\begin{figure}[b]
\centerline{\includegraphics[width=10cm]{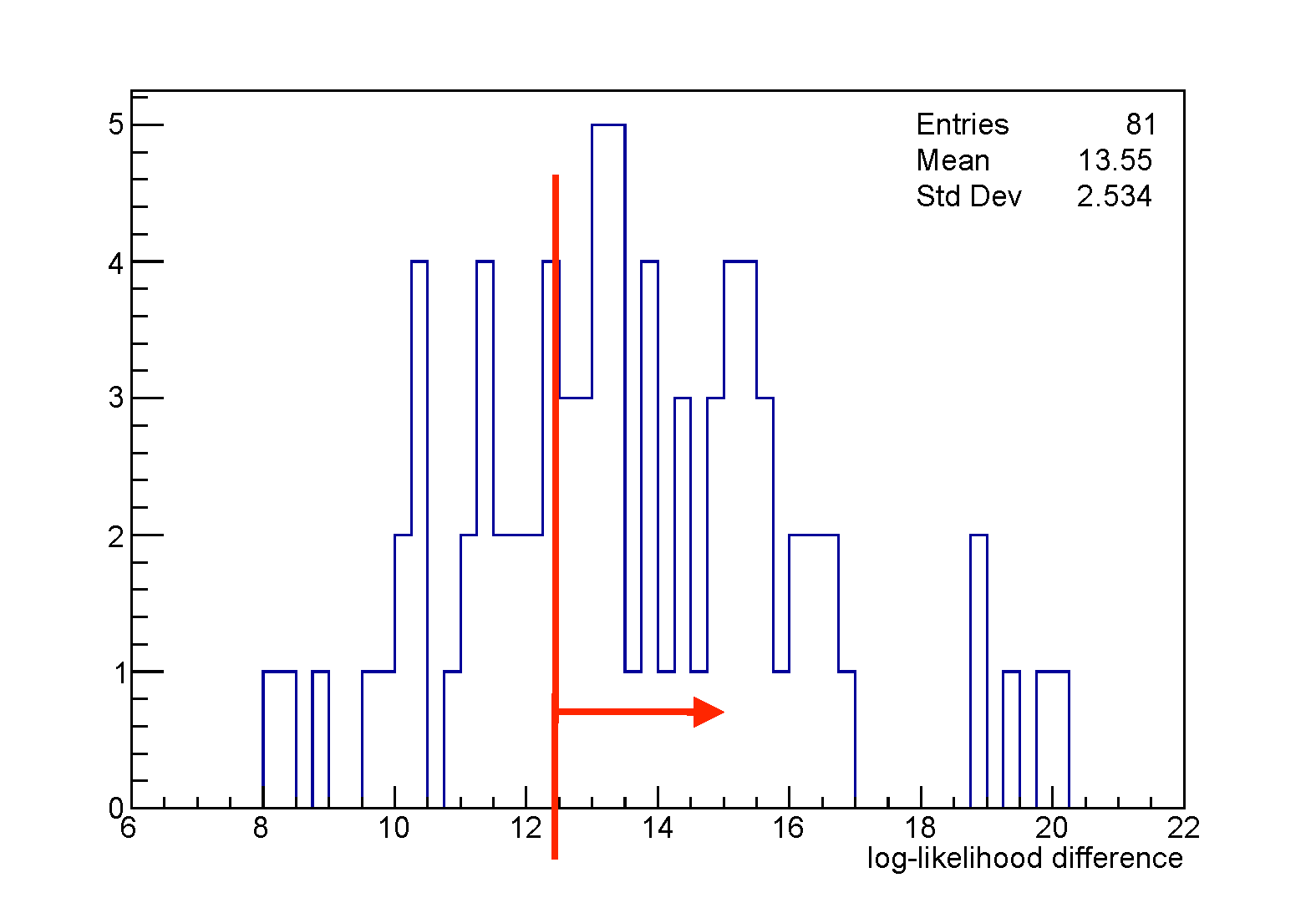}}
\caption{Log-likelihood difference for combined samples with 50$\%$ additional signal and 50$\%$ additional background.  \label{f4}}
\end{figure}

Here we provide a method to estimate the chance of reaching 5$\sigma$ for a 4.1$\sigma$ new signal with a combined data sample which contains $50\%$ additional signal events and 25$\%$, 50$\%$, or 75$\%$ additional background events by simulation.  We generate 50$\%$ of signal events according the signal PDF function, and 25$\%$, 50$\%$, and 75$\%$ of background events according to background PDF function, then combine these simulated events with the original data to evaluate its local significance using the likelihood method described above. We repeated roughly 100 times for each case to see the chance to reach 5$\sigma$.  Figure \ref{f4} shows the log-likelihood ratio for the new sample with 50$\%$ of additional signal and 50$\%$ of additional background, which shows that 54 out 81 cases (66$\%$) reach 5$\sigma$. While the chance is about 39$\%$ and 92$\%$ if the additional background combined is 75$\%$ and 25$\%$, respectively.   

These tests show that certain factors can alter the chance to reach 5$\sigma$ by adding 50$\%$ of additional data, and we can estimate it by using simulated events.

\section{Summary}
As an example of calculating significance, the global significance of a potential new particle has been evaluated with 13409 simulated events in this study. The global significance depends on the assumed mass window and width range defined as a prior. The actual choice for the search window was based on considerations that go beyond the scope of this study.

The additional data added using simulations gave us a better understanding of the chance of reaching a local significance of 5$\sigma$ with different percentages of background events included.

\section*{Acknowledgments}
This work is done by a high school student Kelly J Yi, under guidance of Leonard G Spiegel and Zhen Hu. 


\end{document}